\documentclass[a4paper,11pt,oneside]{article}

\usepackage{amsmath,amssymb,graphicx}

\usepackage{subfig}
\usepackage{mathrsfs}
\usepackage{amsmath}
\usepackage{amssymb}
\usepackage{color}

\usepackage[para]{footmisc} % si no lo ubico antes de {hyperref} surgen problemas con los links de las footnote
\usepackage{xcolor} %para escribir colores por su nombre, pero no me funcionó en las hyperref
\usepackage{hyperref}
\usepackage{graphicx}
\usepackage{tabularx}
\usepackage[latin1]{inputenc}%Toma acentos%
\usepackage{enumerate}
\usepackage{graphicx,amssymb,amstext,amsmath}
\usepackage[bottom=2.0cm,top=2.0cm, textwidth=17cm]{geometry} %textheight=25cm,
\usepackage{cancel}

\usepackage[font={footnotesize}]{caption}

\usepackage{lipsum}
\let\OLDthebibliography\thebibliography
\renewcommand\thebibliography[1]{
  \OLDthebibliography{#1}
  \setlength{\parskip}{0pt}
  \setlength{\itemsep}{8pt plus 0.1ex}
}

\def\ai{\'{\i}}

\def\lp{\left(}
\def\rp{\right)}
\def\lbr{\left[}
\def\rbr{\right]}

\def\be{\begin{equation}}
\def\ee{\end{equation}}
\def\ba{\begin{array}}
\def\ea{\end{array}}
\def\bea{\begin{eqnarray}}
\def\eea{\end{eqnarray}}

\def\inf{\infty}
\def\t{\theta}
\def\T{\Theta}
\def\lb{\label}
\def\d{\delta}

\def\pd{\partial}
\def\r{\rho}

\def\n{\nu}
\def\a{\alpha}
\def\b{\beta}
\def\g{\gamma}
\def\s{\sigma}
\def\Om{\Omega}
\def\om{\omega}
\def\l{\lambda}
\def\P{\Phi}
\def\u{\upsilon}

\begin{document}

\title{\vspace{-1.5cm} \bf Effect of a cylindrical thin-shell of matter on the electrostatic self-force on a charge}
\author{E. Rub\ai n de Celis\footnote{e-mail: erdec@df.uba.ar}\\
{\footnotesize Departamento de F\ai sica, Facultad de Ciencias Exactas y Naturales, Universidad de}\\
{\footnotesize Buenos Aires and IFIBA, CONICET, Cuidad Universitaria, Buenos Aires 1428, Argentina.}}
\date{\small \today}

\maketitle
\vspace{-0.6cm} 
\begin{abstract}
The electrostatic self-force on a point charge in cylindrical thin-shell space-times is interpreted as the sum of a \textit{bulk} field and a \textit{shell} field. The \textit{bulk} part corresponds to a field sourced by the test charge placed in a space-time without the shell. The \textit{shell} field accounts for the discontinuity of the extrinsic curvature ${\kappa^p}_q$. An equivalent electric problem is stated, in which the effect of the shell of matter on the field is reconstructed with the electric potential produced by a non-gravitating charge distribution of total image charge $Q$, to interpret the shell field in both the interior and exterior regions of the space-time. The self-force on a point charge $q$ in a locally flat geometry with a cylindrical thin-shell of matter is calculated. The charge is repelled from the shell if ${\kappa^{p}}_{p}=\kappa<0$ (ordinary matter) and attracted toward the shell if $\kappa>0$ (exotic matter). The total image charge is zero for exterior problems, while for interior problems $Q/q=-\kappa \, r_e$, with $r_e$ the external radius of the shell. The procedure is general and can be applied to interpret self-forces in other space-times with shells, e.g., for locally flat wormholes we found $Q_{\mp}^{wh}/q=-1/ (\kappa_{wh} r_{\pm})$.\\

\noindent 
PACS number(s): 4.20.-q, 04.20.Gz, 04.40.-b, 41.20.Cv\\
% 4.20.-q Classical General Relativity,  04.20.Cv Fundamental problems and general formalism, 04.20.Gz Spacetime topology, causal structure, spinor structure, 04.20.Jb Exact Solutions, 04.40.-b Self-gravitating systems; continuous media and classical fields in curved spacetime, 41.20.Cv Electrostatics; Poisson and Laplace equations, boundary-value problems,11.27.+d Extended classical solutions; cosmic strings, domain walls, texture
Keywords: General relativity; electrostatic self force; thin shells; cylindrical spacetimes; wormholes

\end{abstract}

\section{Introduction}

The problem of a point charge in a curved background is a testing probe for studying the consequences of curvature on electrodynamics. The Maxwell equations in curved space-times for a freely falling observer are locally identical to those in Minkowski space-time, however, curvature and topology produce global differences. The correct solutions of the equations in curved space-time give rise to the so-called self-interaction of charged particles as their own electromagnetic field is bended by geometry \cite{ppv}. In this work we aim to extend the understanding on how curvature affects this self-interaction considering an electric charge at rest in a space-time (ST) with a cylindrical thin-shell (TS) of matter. The analysis is focused on how the TS modifies the expected field of the charge.

The first case considered of a self-force of physical interest was a static charge in a Schwarzschild metric; it was shown that the self-force on a point charge $q$ is repulsive from the centre and has the form $f\sim\frac{Mq^2}{r^3}$, where $M$ is the mass of the spherical source and $r$ is the usual Schwarzschild radial coordinate of the probe charge. This result was first obtained within linearized general relativity \cite{vil79}, and was later confirmed  working in the framework of the full theory; see \cite{will80} and also \cite{varios}. After this work, the study of the self-interaction of a charge was extended to many backgrounds. Recently, the Schwarzschild geometry with de Sitter and anti de Sitter asymptotics \cite{kuchar}, and black holes in other dimensions have been considered \cite{yago}. Alternatively, the central object has been replaced and the self-force was analyzed as a function of the internal structure and composition of the spherically symmetric central body \cite{2}.

Particularly, the first cylindrically symmetric background used to calculate the self-force on a charge was the locally flat geometry of a straight cosmic string and was done by Linet in \cite{lin}. The force points outwards and has the form $f\sim\frac{\mu q^2}{r^2}$, where $\mu$ is the mass per unit length of the string and $r$ is the distance from the string to the charge. This non-vanishing force, associated to the deficit angle of the geometry, illustrates how the global properties of a manifold are manifested by electrostatics. 

The problem of a charge in a TSST was specially addressed in the context of wormholes constructed with shells of exotic matter. Spherically symmetric wormholes were considered in Refs. \cite{whs}. In a previous work we studied the case of a charge in the Schwarzschild wormhole. We found that the self-force differs from that in the black hole space-time, probing the different topology of locally indistinguishable geometries \cite{eoc1}. On the other hand, cylindrical TS wormholes, studied extensively in Refs. \cite{ctsw}, have also been considered to evaluate the self-force problem. In \cite{eoc} we showed that the non trivial topology of locally flat wormholes contribute with an attractive force toward the throat. 

In this work we aim to study the consequences over the electrostatic field and self-force on a test charge produced in cylindrical TSST constructed with positive or negative concavity $\kappa$ (trace of the jump on the extrinsic curvature). The shell of circular profile has positive or negative surface energy density and pastes together two different geometries. The main objective is to analyze the electric effect produced by the presence of the shell isolating a so-called \textit{shell} field. To interpret this part of the field we state an equivalent problem, in which the electric effect produced by the shell of matter is replicated by a non-gravitating shell of surface charge density. The total electric field potential is separated in two parts: an inhomogeneous field produced in the \textit{bulk} geometry where the charge is placed (which has to be renormalized) and the \textit{shell} field which includes every other non local contribution generated by the discontinuity of the extrinsic curvature at the shell and the topology of the constituent geometries. Throughout the work we consider an almost everywhere flat space-time with a cylindrical thin-shell of matter to apply this procedure and analyze the self-force problem.

To calculate the self-interaction of a static point charge the divergent electrostatic potential at the position of the charge has to be renormalized. A singular part, which is known to depend only on the local properties of the geometry in a neighborhood of the charge's position, must be removed and a regular homogeneous solution is obtained. Several regularization procedures have been used in the past in order to remove singular potentials. The renormalization method that enjoys the best justification is that of Detweiler and Whiting, which is based in a four-dimensional singular Green function \cite{DW}. Recently in \cite{Casals}, the authors showed the equivalence between this procedure in the case of a static particle in a static ST and Hadamard's two-point function in three dimensions for the computation of electrostatic self-forces. An alternative approach to renormalization, which is also suitable for charged particles at rest in general static curved ST, is the DeWitt-Schwinger asymptotic expansion of the three-dimensional Green function, which was considered recently in \cite{kpl}. In either of these formalisms the renormalized potentials coincide.

The article is organized in the following sections: in Sect. \ref{s2} we make general remarks for the construction of a cylindrical TSST, and in Sect. \ref{s2.1} a locally flat geometry with a thin-shell is considered. In Sect. \ref{s3} the electrostatic field potential in cylindrical coordinates for TSST, using separation of variables, is generically stated and in Sect. \ref{s3.1} it is solved for the space-time in consideration. In Sect. \ref{s4} the renormalization procedure is applied to obtain the regular potential field over the position of a charge in a TSST, in Sect. \ref{s4.1} the \textit{bulk} and \textit{shell} fields are defined, and in Sect. \ref{s4.2} the regular field is obtained in locally flat cases. In Sect. \ref{s5} the \textit{shell} field is interpreted defining an equivalent problem (Sect. \ref{s5.1}), in Sect. \ref{s5.2} the self-force on a charge in an arbitrary position on the locally flat TSST is computed, analyzed and showed in a numerical plot. In Sect. \ref{s5.3} a complementary example in a wormhole is given. See Sect. \ref{s6} for a summary of the results.

The geometrized unit system is used where $c=G=1$. Throughout the article cylindrical coordinate systems are used. General space-time events are denoted by $x=(t,\bf x)$, with spatial position ${\bf x}=(r, \t, z)$, and $t\in\lp-\inf,+\inf\rp$, $r\in\lp0,+\infty\rp$, $\t\in\lbr0,2\pi\rbr$, $z\in \lp -\inf,+\inf \rp$. Indices with Greek letters $\a$, $\b$, $\g$ are used for four-vector and tensor components which run over $\{t,r,\t,z\}$ where $t$ is the temporal component. The collective index $j=\{r,\t,z\}$ is used exclusively for spatial components. Greek letters also appear with other uses and meanings; these are specified in each case.

\section{The cylindrical thin-shell geometry}
\label{s2}

The Darmois-Israel formalism, found in recent literature \cite{libroVisser} and \cite{Poisson}, is used to construct a static cylindrically symmetric space-time with a thin-shell using coordinates $x=(t,r,\t,z)$ which are continuous in a neighborhood of the shell. To this end, a complete manifold $\Omega_{1}=\{ x_1: \, r_1 \in \lp 0,+\inf \rp \}$, with given metric components $[g_1(r_1)]_{\a \b}$, is cut at the time-like hypersurface $\partial \Omega_{int}=\{r_1 = r_i \}$ to use the four-dimensional interior region $\Omega_{int}=\{ x_1: \, r_1 \in \lp 0,r_i \rbr \}$. A second complete manifold $\Omega_{2}=\{ x_2: \, r_2 \in \lp 0,+\inf \rp \}$ with $[g_2(r_2)]_{\a \b}$ is cut at the hypersurface $\partial \Omega_{ext}=\{ r_2 = r_e \}$ to use the exterior region $\Omega_{ext}=\{ x_2: \, r_2 \in \lbr r_e,+\inf \rp \}$, and paste $\Om_{int}$ with $\Om_{ext}$ at their boundaries: $\partial \Omega_{int} \equiv \partial \Omega_{ext}=H$. The resulting space-time is a complete manifold $\Om=\Om_{int} \cup \Om_{ext}$ such that
\be
\Om = \{ x\} \equiv \left\{ \ba{ll}
\Om_{int}=\{ (t,\r(r),\t,z)\},\; \mbox{if}\; r\in\lp 0,r_i \rbr \,,\\
\Om_{ext}=\{ (t,\r(r),\t,z)\},\; \mbox{if}\; r\in\lbr r_i,+\inf \rp \,,
\ea \right.
\ee
\be \lb{ro}
\mbox{with} \; \r(r)\dot{=}\left\{ \ba{ll}
r_1 = r \,, \; \mbox{if}\; r \in \lp 0,r_i \rbr \,,\\
r_2 = r-r_i+r_e \,,\; \mbox{if}\; r \in \lbr r_i,+\inf \rp \,.
\ea \right.
\ee
Two coordinate systems have been introduced to describe $\Om$; the hypothetically global coordinates $x$, or the patches of $x_1$ and $x_2$ for the respective regions $\Om_{int}$ and $\Om_{ext}$. The geometry has a cylindrical thin-shell at hypersurface $H$ at $r=r_i$ (or $\{r_1=r_i\} \equiv \{r_2=r_e\}$).
The first fundamental form (junction condition) establishes the continuity of the induced metric at $H$ in some coordinates $y^{p}=(t,\t,z)$ installed on both sides of the hypersurface. On the other hand, the second fundamental form
\be
K^{(a)}_{pq} = n_{\a;\b} \, \frac{\pd x_a^{\a}}{\pd y^{p}} \frac{\pd x_a^{\b}}{\pd y^{q}}\, ,
\ee
where $n_{\a}$ is the normal to the shell (pointing from $\Om_{int}$ to $\Om_{ext}$) and $\pd x_a^{\a}/\pd y^{p}$ are the tangent vectors to $H$ with $a=1$ or $2$, is used to compute the discontinuity of the extrinsic curvature $\kappa_{pq}=K_{pq}^{(2)}-K_{pq}^{(1)}$, and its trace at $r=r_i$:
\be
\kappa \equiv {\kappa^{p}}_{p} = {n^{\a}}_{;\a}\,.
\ee
Notice $\kappa$ is equal to the expansion of a congruence of geodesics orthogonal to the shell \cite{Poisson}, i.e., the thin-shell is said to be convex if $\kappa>0$ and concave if $\kappa<0$. Finally, the surface stress-energy,
\be
{S^{p}}_{q}=-\frac{1}{8 \pi}({\kappa^{p}}_{q}-\kappa \, {\delta^p}_q)\,,
\ee
determines the surface energy density $\Sigma=-{S^t}_t$ and pressures $P_{\t}={S^{\t}}_{\t}$, $P_{z}={S^{z}}_{z}$ at the shell.

\subsection{Locally flat space-times with a cylindrical thin-shell of matter}
\lb{s2.1}

We will consider a space-time with line element
\be\lb{tsg}
ds^2 = -dt^2 + dr^2 + \lbr\rho_{\om}(r)\rbr^2 d\t^2 + dz^2,
\ee
with profile function
\be
\rho_{\om}(r) = \T(r_i - r) \om_i r + \T(r - r_i) \om_e\lbr r - r_i + r_e \rbr\,,
\ee
in global coordinates $x=(t,r,\t,z)$. Parameters $\om_i, \om_e, r_i, r_e$ arise from the construction of this geometry using two infinitely thin gauge cosmic string manifolds of different deficit angles, $2\pi(1-\om_i)$ and $2\pi(1-\om_e)$, and line elements,
\be\lb{cs1}
ds^2_{(1)}= -dt^2 + dr_1^2 + \om_i^2 r_1^2 d\theta^2 + dz^2\,,
\ee
\be\lb{cs2}
ds^2_{(2)}= -dt^2 + dr_2^2 + \om_e^2 r_2^2 d\theta^2 + dz^2,
\ee
respectively. The magnitudes $\om_i, \om_e \in (0;1]$ are related to the mass per unit length $\mu=\frac{1-\om}{4}$ of the string, so the energy-momentum tensor is $T_\a^\b=\rm{diag}(\mu,0,0,\mu)\,\delta(r)/(2 \pi \om r)$ in the cosmic string ST \cite{libroVilenkin}. Consequently space-time (\ref{tsg}) is locally flat except at the cylindrical thin-shell at $r=r_i$ ($\om_i \neq \om_e$) or at the conical singularity at $r=0$ ($\om_i \neq 1$), and has a conical asymptotic region ($\om_e \neq 1$). The induced metric at the shell has line element
\be
dH^2 = - dt^2 + h_{\t \t} d\t^2 +dz^2\,,
\ee
and the junction condition determines
\be
\sqrt{h_{\t \t}}=\om_i r_i=\om_e r_e\,.
\ee
The associated extrinsic curvature from each side of the shell has the only non-vanishing components: $K^{(2)}_{\t\t}=\om_e^2 r_e$ and $K^{(1)}_{\t\t}=\om_i^2 r_i$, at $r=r_i^{\pm}$, respectively. The trace of the discontinuity $\kappa_{pq}$ is
\be
\kappa=\frac{\om_e-\om_i}{\om_ir_i}\,,
\ee
and the surface energy density at the shell for the locally flat cylindrical space-time is
\be
\Sigma= -\frac{\om_e-\om_i}{8 \pi \om_ir_i}\,.
\ee
We see the shell is concave and made of ordinary matter if $\om_i>\om_e$, or convex and made of exotic matter if $\om_e>\om_i$.

\section{Electrostatic field potential}
\lb{s3}

The calculation of the self-interaction of a charge $q$ starts from Maxwell equations for curved space-time:
\begin{eqnarray}\lb{max}
\left(F^{\b\a}\sqrt{-g}\right)_{,\b} & = & 4\pi j^\a \sqrt{-g}\,,\\
F_{\a\b} & = & A_{\b,\a}-A_{\a,\b}\,,\\
j^\a & = & \frac{q}{\sqrt{-g}}\delta({\bf x},{\bf x}')\frac{dx^\a}{dt}\,,
\end{eqnarray}
where $g$ is the metric determinant, $A_\a$ is the four-potential, and $j^\a$ is the four-current. For a static diagonal cylindrically symmetric $g_{\a \b}$ and a rest charge at ${\bf x}'=(r',\t',z')$ we have
\be\lb{Pgen}
 \lp g^{tt}\, g^{jj}\, \P,_{j} \sqrt{-g} \rp,_{j}=4\pi q \, \delta({\bf x},{\bf x}') \, ,
\ee
for the component $A_t \equiv \P$ of the four-potential, while the other $j$ components vanish, with $\delta({\bf x},{\bf x}')=\d(r-r')\d(\t-\t')\d(z-z')$.
To solve the electrostatic problem the following expansion for the potential will be used:
\be\lb{exp}
\P=q \int \limits_{0}^{+\inf}dk\,Z(k,z) \sum_{n=0}^{+\inf}Q_n(\t) R_{n,k}(r)  \, ,
\ee
where $F_z=\{Z(k,z)=\cos[k(z-z')]\}$ and $F_{\t}=\{ Q_n(\t)=a_n \cos[n(\t-\t')]\}$ are a complete set of orthogonal functions of the coordinates $z$ and $\t$ of the cylindrical set $(t,r,\t,z)$, with
\be
a_n = \left\{ \ba{ll}
         \frac{2}{\pi} & \mbox{if $n=0$}\,, \\
         \frac{4}{\pi} & \mbox{if $n \, \epsilon \, \mathbb{N}-\left\{0 \right\}$}\,,
         \ea \right.
\ee
and $R_{n,k}(r)$ radial functions satisfying
\be\lb{ecrad}
\lbr \frac{\pd}{\pd r} \lp \frac{\sqrt{-g}}{g_{tt}\, g_{rr}} \, \frac{\pd}{\pd r} \rp - \frac{\sqrt{-g}}{g_{tt}} \lp \frac{n^2}{g_{\t \t}} + \frac{k^2}{g_{zz}} \rp \rbr R_{n,k}(r) = \delta(r-r') \, ,
\ee
for each $n$ and $k$. In a TSST, the metric is different in each side of the shell so the radial function solutions will be	 
\be  \lb{rad}
R^{(a)}_{n,k}(r)= \left\{ \ba{ll}
R^{(1)}_{n,k}(r_1)\quad\mbox{for $\Om_{int}$}\,,\\
R^{(2)}_{n,k}(r_2)\quad \mbox{for $\Om_{ext}$}\,, 
\ea \right.
\ee
where the global radial coordinate $r$ is replaced by $r_1$ or $r_2$, using (\ref{ro}). $R^{(1)}_{n,k}(r_1)$ and $R^{(2)}_{n,k}(r_2)$ are general solutions to the corresponding homogeneous version of (\ref{ecrad}), and the boundary conditions are stated in terms of these functions. Finiteness at the origin and at infinity is expressed by
\be \lb{0}
\lim_{r_1 \rightarrow 0} R^{(1)}_{n,k} \neq \inf \,,
\ee
\be \lb{inf}
\lim_{r_2 \rightarrow +\inf} R^{(2)}_{n,k} = 0\,.
\ee
To paste together solutions in each region we invoke the homogeneous Maxwell equations $F_{[\a\b;\g]}=0$, which guarantee the continuity of the tangential component of the electric field, and require $\P$ to satisfy the usual continuity condition at the shell:
\be \lb{cont}
R^{(1)}_{n,k} \biggl|_{r_1=r_i} = R^{(2)}_{n,k} \biggl|_{r_2=r_e}.
\ee
On the other hand, integrating (\ref{ecrad}) in the radial coordinate $r$ over an interval $[r_0-\epsilon;r_0+\epsilon]$ with $\epsilon \rightarrow 0$, $r_0=r_i$, and the charge placed at $r'\neq r_i$, we obtain 
\be \lb{jump}
\lp \frac{\pd}{\pd r_2} R^{(2)}_{n,k} \rp \biggl|_{r_e^+} - \lp \frac{\pd}{\pd r_1} R^{(1)}_{n,k}\rp \biggl|_{r_i^-} = 0 \,,
\ee
where we argue convergence for $\P$ and the fact that $g^{(a)}_{\a \b}$, $R^{(a)}_{n,k}$ and $\pd_r R^{(a)}_{n,k}$ do not diverge at $r_i^{\pm}$.\footnote{Note that the factor $\pd_r (\frac{\sqrt{-g}}{g_{tt}\, g_{rr}})$ in Eq. (\ref{ecrad}) generates a delta function due to the step-jump of the metric at $r=r_i$ but it does not appear in (\ref{jump}), i.e., the extra terms arising from the discontinuity of the extrinsic curvature over the shell do not introduce electric sources to the electrostatic potential.}
Finally we separate two cases: (I) $r'<r_i$ ($r'_1$ in $\Om_{int}$), or (II) $r'>r_i$ ($r'_2$ in $\Om_{ext}$). The submanifold where the charge is placed is divided in two regions and we establish continuity of $\P$ and the jump of the electric field to paste solutions from each side of $r'$: 
\be\lb{cont2}
R^{(a)}_{n,k} \big|_{{r'_{a}}^{+}} = R^{(a)}_{n,k} \big|_{{r'_{a}}^{-}}\, ,
\ee
\be \lb{jump2}
 \lp \frac{\pd}{\pd r_a} R^{(a)}_{n,k} \rp \biggl|^{{r'_{a}}^{+}}_{{r'_{a}}^{-}} = \frac{g^{(a)}_{tt}\,g_{rr}^{(a)} }{\sqrt{-g_{(a)}}} \biggl|_{r'_{a}}\,, 
\ee
with $a=1$ and $r'_1=\r(r')$ in case (I), or $a=2$ and $r'_2=\r(r')$ in case (II). To obtain (\ref{jump2}) we operate analogously as for (\ref{jump}) with $r_0=r'$.

\subsection{Potential field in locally flat space-times with a cylindrical thin-shell}
\lb{s3.1}

The Poisson equation (\ref{Pgen}) for a point-like charge $q$ at rest in ${\bf x}'=(r',\t',z')$ of metric (\ref{tsg}), where $-g_{tt}=g_{rr}=g_{zz}=1$ and $g_{\t \t}=-g=\r_{\om}^2$, reads
\be\lb{p}  
\r_{\om} \lbr \frac{\pd^2}{\pd r^2} + \frac{1}{\r_{\om}} \lp \frac{\pd \r_{\om}}{\pd r} \rp \frac{\pd}{\pd r} + \frac{1}{\r_\om^2} \frac{\pd^2}{\pd \t^2} + \frac{\pd^2}{\pd z^2} \rbr \P = - 4\pi q \,\delta({\bf x},{\bf x}')\,.
\ee
To solve the potential $\P$ in this locally flat TSST it is easier to show first the solutions $\P^{CS}_{\om}$ and $\P^M$ for a point charge in the cosmic string and Minkowski geometries, respectively. The gauge cosmic string ST is described by any of both metrics (\ref{cs1}) or (\ref{cs2}), or equivalently (\ref{tsg}) with $\om_i=\om_e=\om$. The profile function becomes $\r_{\om}(r)=\om r$, while for Minkowski ST the potential $\P^M$ corresponds to the solution of (\ref{p}) with $\om=1$. Using the expansion (\ref{exp}), Eq. (\ref{ecrad}) for $R^*_{n,k}$ corresponds to the $\upsilon {th}$ order inhomogeneous modified Bessel equation,
\be\lb{bessel}
\left[\frac{\pd^2}{\pd r^2}+\frac{1}{r}\frac{\pd}{\pd r}-\left(k^2+\frac{\upsilon^2}{r^2}\right)\right] R^{*}_{n,k} =
-\frac{\d(r-r')}{\om r}\,,
\ee
where $\upsilon=n/\om$. Requiring finiteness at $r\rightarrow 0$ and at $r\rightarrow\infty$, the continuity at $r=r'$ and its derivative discontinuity determined from (\ref{bessel}) we obtain	
\be \lb{csrf}
R^{*}_{n,k} = \frac{1}{\om} K_{\upsilon}(kr_>)I_{\upsilon}(kr_<) =
\left\{\ba{ll}
				R^{M}_{n,k} & \mbox{if $\om=1$}\,, \\
				R^{CS}_{n,k} & \mbox{if $0<\om<1$}\,, \\
				\ea \right.
\ee
where $K_{\upsilon}(kr)$ and $I_{\upsilon}(kr)$ are the usual modified Bessel functions of order $\upsilon=n/\om$ with $n \, \epsilon \, \mathbb{N}_0$, two independent solutions of the homogeneous version of Eq. (\ref{bessel}) and $r_{\stackrel{>}{<}}=\left\{ r ; r' \right\}$.

Finally for (\ref{tsg}), which has an interior region $\Om_{int}$ of deficit angle $2\pi(1-\om_i)$ and an exterior $\Om_{ext}$ of deficit angle $2\pi(1-\om_e)$ separated by the cylindrical thin-shell at $r=r_i$, the radial functions will have general solutions:
\be 
R^{(1)}_{n,k} = C_1 I_{\l}(k r_1) + C_2 K_{\l}(k r_1)  \, , \quad  \mbox{for $\Om_{int}$} \,,
\ee
\be
R^{(2)}_{n,k} = C_3 I_{\n}(k r_2) + C_4 K_{\n}(k r_2) \, , \quad\mbox{for $\Om_{ext}$} \,,
\ee
where $\l=n/\om_i$, $\n=n/\om_e$, and coefficients $C$ are functions of $n$ and $k$ to be determined with the boundary conditions stated in Sect. \ref{s3}. The possible solutions are:
\\
(I). Charge in region $\Om_{int}$ ($r'<r_i$): $r'_1=\r(r')$  
\be\lb{r-}\ba{ll}
R^I_{n,k} =
\left\{
\ba{ll}
\frac{1}{\om_i} I_{\l}(kr_{1<}) \lbr A_{n,k} I_{\l}(kr_{1>}) + K_{\l}(kr_{1>}) \rbr \, ,& \, \mbox{for $\Om_{int}$}\,, \\
\frac{1}{\om_i} \frac{I_{\l}(kr'_1)}{K_{\n}(k r_e)} \lbr A_{n,k} I_{\l}(kr_i) + K_{\l}(kr_i) \rbr K_{\n}(k r_2) \, ,& \, \mbox{for $\Om_{ext}$}\,, \\
\ea
\right.\\
\mbox{$r_{1>} =\,$ max $\left\{ r_1; r'_1 \right\}$,  $r_{1<} =\,$ min $\left\{ r_1; r'_1 \right\}$}\,, \\\\
A_{n,k} = \frac{K_{\n}(k r_e) \lbr \pd_r K_{\l}(kr) \rbr |_{r_i}  - K_{\l}(k r_i) \lbr \pd_r K_{\n}(k r) \rbr |_{r_e}}{I_{\l}(k r_i) \lbr \pd_r K_{\n}(k r) \rbr |_{r_e} - K_{\n}(k r_e) \lbr \pd_r I_{\l}(k r) \rbr |_{r_i}} \,.
\ea
\ee
(II). Charge in region $\Om_{ext}$ ($r'>r_i$): $r'_2=\r(r')$
\be\lb{r+}\ba{ll}
R^{II}_{n,k} =
\left\{
\ba{ll}
\frac{1}{\om_e} \lbr I_{\n}(k r_e) + B_{n,k} K_{\n}(k r_e) \rbr \frac{K_{\n}(k r'_2)}{I_{\l}(kr_i)} I_{\l}(kr_1)  \, , \, & \mbox{for $\Om_{int}$} \,,\\
\frac{1}{\om_e} \lbr I_{\n}(k r_{2<}) + B_{n,k} K_{\n}(k r_{2<}) \rbr K_{\n}(k r_{2>}) \, , &\, \mbox{for $\Om_{ext}$}\,, \\
\ea
\right. \\
\mbox{$r_{2>} =\,$ max $\left\{ r_2; r'_2 \right\}$,  $r_{2<} =\,$ min $\left\{ r_2; r'_2 \right\}$}\,, \\\\
B_{n,k} = \frac{I_{\n}(k r_e) \lbr \pd_r I_{\l}(kr) \rbr |_{r_i}  - I_{\l}(k r_i) \lbr \pd_r I_{\n}(k r) \rbr |_{r_e}}{I_{\l}(k r_i) \lbr \pd_r K_{\n}(k r) \rbr |_{r_e} - K_{\n}(k r_e) \lbr \pd_r I_{\l}(k r) \rbr |_{r_i}}\,.
\ea
\ee

\section{Renormalization procedure}
\lb{s4}

To obtain the self-interaction we have to perform the renormalization of the electrostatic potential. In either of the formalisms mentioned in the introduction the regular potential field is
\be\lb{PR}
\P_{R}(x'^{j})=\lim_{x^{j}\to x'^{j}}(\P-\P_{S})\,,
\ee
where the coincidence limit takes the coordinate spatial components $x^j$ to the charge's position $x'^{j}$ along the shortest geodesic connecting them. In this definition, the singular term is
\be \lb{psing}
\P_{S}=\sqrt{-g_{t't'}}G^{(3)}_S(x^j;x'^{j}),
\ee
with $g_{t't'}=g_{tt}(x')$ and $G^{(3)}_S(x^j,x'^{j})$ the singular Green function in three dimensions \cite{hadamard}. The Green function must have the same singularity structure as the particle's actual field and exert no force on the particle. The three methods mentioned agree in the following expansion \cite{DW,kpl}:
\be\lb{Gs}
G^{(3)}_S(x^j;x'^{j})=\frac{q}{\sqrt{2\g}} \lbr 1- \lp \frac{g_{t't',j'}\g^{,j'}}{4g_{t't'}} \rp  +\mathcal{O}\left(\g\right) \rbr .
\ee
The function $\g=\g(x^j,x'^{j})$ in (\ref{Gs}) is half the squared geodesic distance between $\bf {x}$ and $\bf {x}'$ as measured in the purely spatial sections of the space-time and $\g^{,j}=g^{jl}\partial \g / \partial x^{l}$ (see Refs. \cite{Casals} or \cite{kpl}  for a full derivation\footnote{In (\ref{Gs}) there is an overall sign difference with respect to the cited literature arising from the convention taken for the Maxwell equations (\ref{max}).}). The terms of order $\mathcal{O}\left(\g\right)/\sqrt{2\g}$ are irrelevant for the renormalization of the potential field since they vanish in the coincidence limit (\ref{PR}). 

\subsection{Regular field in a space-time with a thin-shell}
\lb{s4.1}

To take the renormalization procedure (\ref{PR}) we concentrate on the solution $\P$ in a neighborhood of the charge placed in the four-dimensional region $\Om_{int}$ or $\Om_{ext}$. From the explicit results in Sect.  \ref{s3.1} in a locally flat TSST we notice that, in general, when the observation point and the test charge are in the same region $\Om_{int}$ ($\Om_{ext}$), the potential field can be read as two terms:
\be\lb{terms}
\Phi=\Phi^{bulk}+\Phi^{\sigma}\,, \quad \mbox{if $x, x' \in \Om_{int}\,(\Om_{ext})$}\,.
\ee
The \textit{bulk} part $\P^{bulk}$ is the field sourced by the electric charge in $\Om_{int}$ ($\Om_{ext}$), it is completely defined by extending its domain to the complete manifold $\Om_{a}=\{ x_a=(t,r_{a},\t,z) \}$ with $a=1$($2$), such that:
\be\lb{defbulk}
\lp g_{(a)}^{tt}\, g_{(a)}^{jj}  (\P^{bulk}),_{j} \sqrt{-g_{(a)}} \rp,_{j} = 4\pi q \delta({\bf x}_a,{\bf x}'_a) \,,\, \forall x_a \in \Om_{a}\,,
\ee
$$
\P^{bulk} \stackrel{|{\bf x}_{a}| \rightarrow \inf}{\longrightarrow} 0 \,, \qquad
\P^{bulk} \stackrel{|{\bf x}_{a}| \rightarrow 0}{\longrightarrow} \mbox{finite} \,,
$$
with ${\bf x}'_a=(r_a'=\r(r'),\t',z')$. The part of the field left over, $\P^{\s}$, will be called the \textit{shell} term and accounts for the presence of the thin-shell of matter joining the two constituent submanifolds $\Om_{int}$ and $\Om_{ext}$. The divergent field at the position of the charge is found in $\P^{bulk}$, consequently, the renormalization procedure yields \footnote{Linearity of the Maxwell equations in a fixed curved space-time guarantees (\ref{reg}) is a regular solution at the position of the charge.}
\be \lb{reg}
\Phi_R(x'^{j})= \Phi^{bulk}_R(x'^{j})+\P^{\sigma}(x'^{j})\,,
\ee
\be
\Phi^{bulk}_R(x'^{j})= \lim_{x^{j}\to x'^{j}}(\P^{bulk}-\P_S)\,. \\
\ee

\subsection{Regularization in locally flat space-times with a cylindrical thin-shell}
\lb{s4.2}

In space-time (\ref{tsg}) the geometry is locally flat in a neighborhood of the charge in either of the cases (I) or (II), (\ref{psing}) rapidly yields $\P_{S}=\P^M$. This is the singular term encountered in any cosmic string geometry (\ref{cs1}) or (\ref{cs2}) because they are locally indistinguishable. Distributing terms in the radial functions (\ref{r-}) or (\ref{r+}) we obtain
\be\lb{bulk}
\Phi^{bulk}= \left\{ \ba{ll}
 \Phi^{CS}_{\om_i}(r_1,\t,z)\,, \quad \mbox{if $x_1, x_1' \in \Om_{int}$}\,, \\
 \Phi^{CS}_{\om_e}(r_2,\t,z)\,, \quad \mbox{if $x_2, x_2' \in \Om_{ext}$}\,,
\ea \right.
\ee
the cosmic string potential solved in (\ref{csrf}) with $r_a'=\r(r')$ in the corresponding case $a=1,2$; and
\be\lb{potsigma}
\Phi^{\sigma} = q \int \limits_{0}^{+\inf}dk\, Z(k,z) \sum_{n=0} ^{+\inf}Q_n(\t) \left\{ \ba{ll}
\frac{A_{n,k}}{\om_i} I_{\l}(k r_1') I_{\l}(k r_1)\,, \quad \mbox{if $x_1, x_1' \in \Om_{int}$}\,, \\
\frac{B_{n,k}}{\om_e} K_{\n}(k r_2') K_{\n}(k r_2)\,, \quad \mbox{if $x_2, x_2' \in \Om_{ext}$}\,.
\ea \right.
\ee
The bulk part of the field is the electrostatic potential of the charge in the respective cosmic string background satisfying definition (\ref{defbulk}) for either case. The shell field $\P^{\s}$ does not diverge at the position of the charge by construction. The renormalization of $\P^{CS}_{\om}$ has already been done by Linet in \cite{lin} and the result is written in a closed form. For $\om=\om_i$ ($\om_e$) we have
\be
\Phi^{bulk}_R= \lim_{x^{j}\to x'^{j}}(\P^{CS}_{\om}-\P^M)= \P^{Linet}_{\om}(r'),
\ee
\be \lb{Lin}
\ba{ll}
\P^{Linet}_{\om}(r')= \frac{q}{2\pi} \frac{L_{\om}}{r'}\, , \\
L_{\om}=\int \limits_{0}^{+\inf} \left[ \frac{\sinh(\zeta/\om)}{\om\left[\cosh(\zeta/\om)-1\right]}-\frac{\sinh{\zeta}}{\cosh{\zeta}-1}\right]\frac{d\zeta}
{\sinh(\zeta/2)} \, .
\ea\ee
It only depends on the radial position $r'$ in virtue of the cylindrical symmetry. Therefore, the self-potential at a general position ${\bf x}'$ of the charge in the thin-shell geometry (\ref{tsg}) is
\be\lb{ren}
\P_{R}(r')=\left\{ \ba{ll}
\frac{q}{2\pi} \frac{L_{\om_i}}{r_1'} + q \int \limits_{0}^{+\inf}dk \sum\limits_{n=0} ^{+\inf} \frac{a_n A_{n,k}}{\om_i} I_{\l}^2(k r_1') \quad \mbox{if $x_1' \in \Om_{int}$}\,,\\
\frac{q}{2\pi} \frac{L_{\om_e}}{r_2'} + q \int \limits_{0}^{+\inf}dk \sum\limits_{n=0} ^{+\inf} \frac{a_n B_{n,k}}{\om_e} K_{\n}^2(k r_2') \quad \mbox{if $x_2' \in \Om_{ext}$}\,.
\ea\right.
\ee

\section{Interpretation of the field and self-force}
\lb{s5}

The term $\P^{\s}$ which appears in $\P_R$ is determined from the construction (\ref{terms}) and the definition of $\P^{bulk}$ and $\P$. This field contains the information on how the thin-shell of matter focuses or defocuses the electric field lines and which effect it produces over the test charge. The shell is uncharged, nevertheless, the discontinuous extrinsic curvature of its hypersurface produces the field lines to bend and interact with the test charge. In the following subsection we give a clear interpretation for $\P^{\s}$.

\subsection{The shell field $\P^{\s}$ and the equivalent problem}
\lb{s5.1}

Equation (\ref{terms}) suggests we can interpret $\P$ with an equivalent electric problem. If the observation point and test charge are in $\Om_{int}$ ($\Om_{ext}$), $\P$ can be thought to be the potential generated by the charge $q$ at $x'_{1(2)}=(t,r'_{1(2)},\t',z')$ in the complete manifold $\Om_{1}$ ($\Om_{2}$) plus an additional field $\P^{\s_{1(2)}}$ sourced by a surface charge layer placed at $\partial \Om_{int}$ ($\partial \Om_{ext}$). Notice that the equivalent problem is stated in the complete manifold $\Om_{1}$ ($\Om_{2}$), one of the primitive space-times used for the construction of the thin-shell geometry, but we are interested in the solution over the region $\Om_{int}$ ($\Om_{ext}$) which appears in the original problem.\footnote{The absence of the gravitating shell in the equivalent problem changes the original geometry on one side of the shell but we will be only concerned on the electric problem over the unchanged side, where the charge is placed.\\} The effect over the electric field potential produced by the shell of matter is replicated with the potential generated by a layer of non-gravitating charge density which produces exactly the same total field $\P^{\s}$ in the region of interest $\Om_{int}$ ($\Om_{ext}$). The equivalent problem has solution
\be \lb{equiv}
\Phi_{a}(r_{a},\t,z)=\Phi^{bulk}_a(r_{a},\t,z)+\Phi^{\sigma_{a}}(r_{a},\t,z)\,,
\ee
$\forall x_{a} \in \Om_{a}$, with $a=1$ if $x, x' \in \Om_{int}$ or $a=2$ if $x, x' \in \Om_{ext}$ in the original problem. $\P_a^{bulk}$  is defined by (\ref{defbulk}). $\P^{\s_a}$ is sourced by a surface charge density $\s_a(\t,z)$ placed at $r_{\s}=\r(r_i^-)=r_i$ if $a=1$, or $r_{\s}=\r(r_i^+)=r_e$ if $a=2$, and this is completely defined by
\be\lb{sigmaa}\ba{ll}
\frac{1}{\sqrt{-g_{(a)}}} \lp g_{(a)}^{tt}\, g_{(a)}^{jj}  (\P^{\s_a}),_{j} \sqrt{-g_{(a)}} \rp_{,j} = 4\pi \delta(r_a - r_{\s}) \s_a(\t,z) \,, \quad \forall x_a \in \Om_{a}\,, \\
\qquad \qquad \qquad \,\,\, \P^{\s_a} \stackrel{|{\bf x}_{a}| \rightarrow \inf}{\longrightarrow} 0 \,,  \qquad
\P^{\s_a} \stackrel{|{\bf x}_{a}| \rightarrow 0}{\longrightarrow} \mbox{finite} \,, \\
\qquad \qquad \qquad \qquad \P^{\s_a}|_{r_{\s}} = \lp \P - \P_a^{bulk}\rp |_{r_a=r_{\s}}\,,
\ea\ee
with
\be \lb{sig}
\s_a(\t,z)=\frac{g_{(a)}^{tt}g_{(a)}^{rr}}{4 \pi} \frac{\pd \P^{\s_a}}{\pd r_a} \biggl|^{r_{\s}^+}_{r_{\s}^-}\,,
\ee
where we used the fact that metrics $g_{(a)}^{\a \b}$ are static and diagonal, and that coordinate base element $\partial r_a$ is perpendicular to the shell.\footnote{The homogeneous field $\P^{\s_{1(2)}}$ over $\Om_{int}$ ($\Om_{ext}$) could have been constructed analogously placing $\s_{1(2)}$ at any finite radius $r_{\s}>\r(r_i^-)$($r_{\s}<\r(r_i^+)$).\\} For $x_1 \in \Om_{int}$, or $x_2 \in \Om_{ext}$, the construction (\ref{equiv}) trivially satisfies
\be\lb{const}
\Phi= \left\{ \ba{ll}
 \P_{1} \lp r_{1},\t,z \rp \,, \quad \mbox{if $x_1, x_1' \in \Om_{int}$}\,, \\
 \P_{2} \lp r_{2},\t,z \rp \,, \quad \mbox{if $x_2, x_2' \in \Om_{ext}$}\,.
\ea \right.
\ee

\subsection{Self-force in locally flat space-times with a cylindrical thin-shell}
\lb{s5.2}

Applying the above procedure to build the equivalent problems in space-time (\ref{tsg}) we have
\be
\P^{bulk}_a= \P^{CS}_{\om}(r_a,\t,z)\,,
\ee
with $\om=\om_i$ ($\om_e$) if $a=1$ ($2$), and
\be\lb{potsig}\ba{ll}
\Phi^{\s_{a}} = q \int \limits_{0}^{+\inf}dk\, Z(k,z) \sum\limits_{n=0} ^{+\inf} Q_n(\t) s_{n,k}^{(a)}
 \left\{\ba{ll}
I_{\l}(k r_<) K_{\l}(k r_>)   \quad \mbox{if $a=1$} \,,\\
I_{\n}(k r_<) K_{\n}(k r_>) \quad \mbox{if $a=2$}\,,
\ea \right. \\
s_{n,k}^{(a)} = 
\left\{\ba{ll}
\frac{A_{n,k}}{\om_i} \frac{I_{\l}(k r_1')}{K_{\l}(k r_i)}\,, \quad \l=\frac{n}{\om_i} \quad \mbox{if $a=1$} \,,\\
\frac{B_{n,k}}{\om_e} \frac{K_{\n}(k r_2')}{I_{\n}(k r_e)}\,, \quad \nu=\frac{n}{\om_e} \quad \mbox{if $a=2$} \, ,
\ea \right. 
\ea  
\ee
where $r_{\stackrel{>}{<}}=\left\{ r_1; r_i \right\}$, $r_1'=\r(r')$ or $r_{\stackrel{>}{<}}=\left\{ r_2; r_e \right\}$, $r_2'=\r(r')$) for $a=1,2$, respectively. These two terms sum up the potential $\P=\P_a^{bulk}+\P^{\s_a}$ in a neighborhood of the charge in $\Om_{int}$ for $a=1$ or $\Om_{ext}$ for $a=2$. We will consider the equivalent problems to interpret the results of the self-force in a locally flat TSST. 

\begin{figure} [h!]
 \centering
  \subfloat[$\om_i=1$ and $\om_e=0.9$.]{
   \label{1a}
    \includegraphics[width=0.45\textwidth]{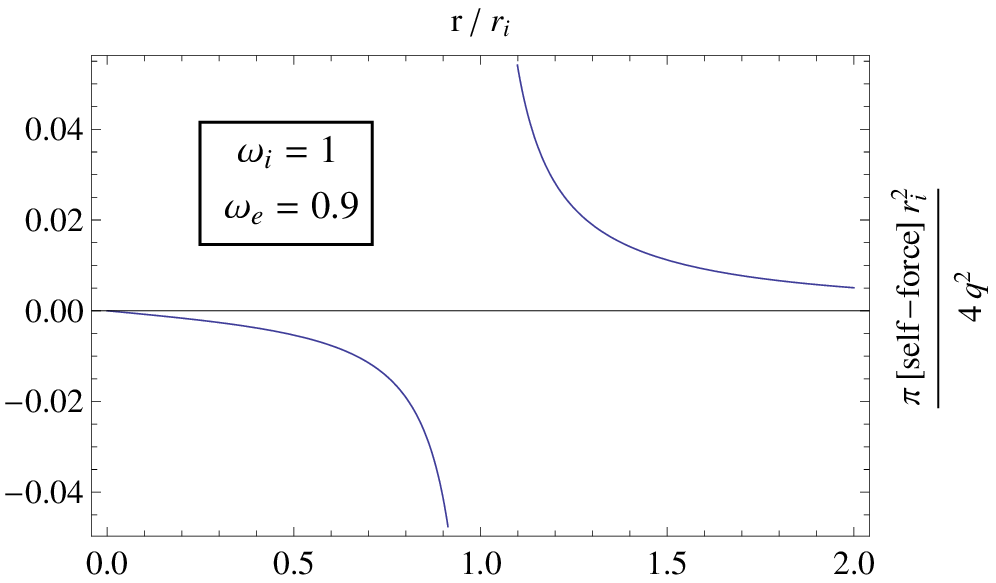}}
  \subfloat[$\om_i=0.9$ and $\om_e=0.7$.]{
   \label{1b}
    \includegraphics[width=0.45\textwidth]{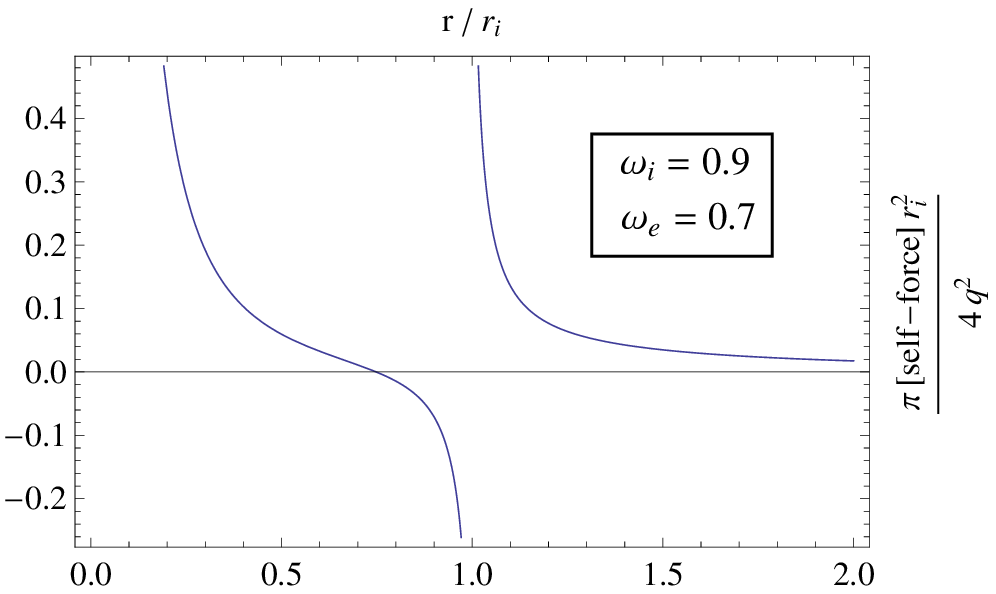}}
 \caption{Dimensionless radial self-force $\frac{\pi r_i^2}{4q^2}\mathrm{f}_{self}$ as a function of $r/r_i$ in a background with a thin-shell made of ordinary matter and $\kappa<0$. Negative values correspond to a force toward the central axis. The shell repels the charge.}
 \label{1}
\end{figure}

\begin{figure} [h!]
 \centering
  \subfloat[$\om_i=0.5$ and $\om_e=1$.]{
   \label{2a}
    \includegraphics[width=0.45\textwidth]{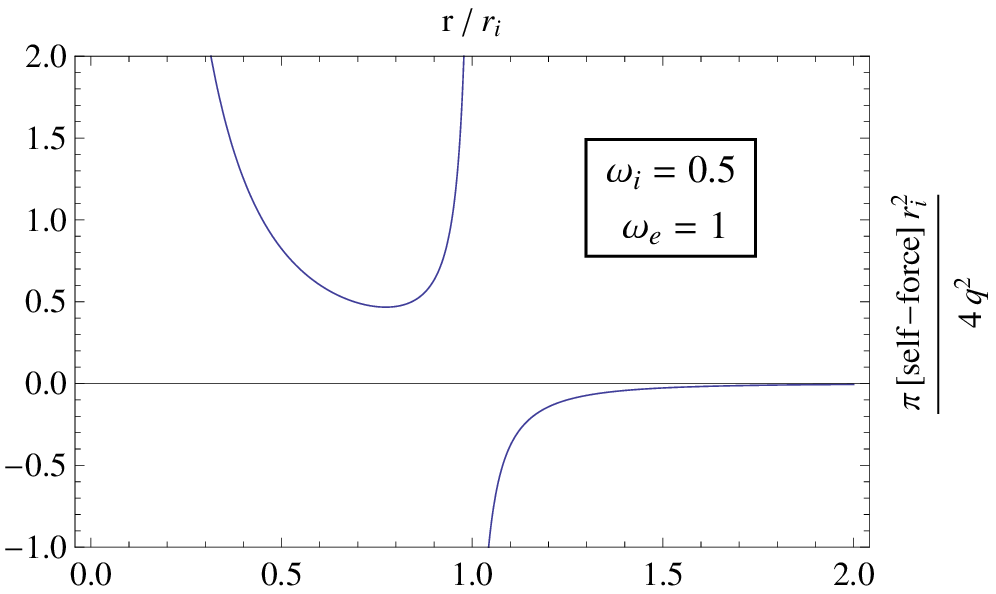}}
  \subfloat[$\om_i=0.1$ and $\om_e=0.3$.]{
   \label{2b}
    \includegraphics[width=0.45\textwidth]{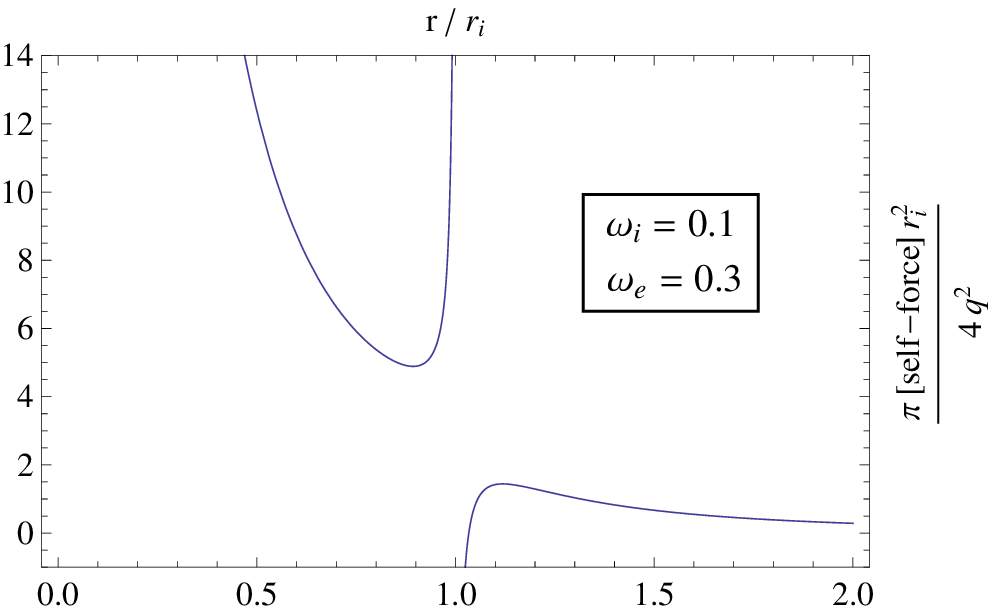}}
 \caption{Dimensionless radial self-force $\frac{\pi r_i^2}{4q^2}\mathrm{f}_{self}$ as a function of $r/r_i$ in a background with a thin-shell made of exotic matter and $\kappa>0$: the shell attracts the charge.}
 \label{2}
\end{figure}

The self-energy is defined as $U_{self}=\frac{q}{2} \P_R(r)$, the electrostatic self-force is radial and is computed as\footnote{This corresponds to the force observed by a static observer at the position of the charge or equivalently to the contravariant tetrad component of the force calculated from $F^{\a\b}$ using the regular potential $\P_R(r,r')$ before taking the coincidence limit, with the final result evaluated at the charge's position; see for example \cite{eoc1} for the explicit derivation.} $\mathrm{f}_{self}=-\pd_r U_{self}$:
\be \lb{f}
\mathrm{f}_{self}=
\left\{\ba{ll}
\frac{q^2}{4\pi} \frac{L_{\om_i}}{r^2} -\frac{q^2}{\om_i} \int \limits_{0}^{+\inf}dk \sum\limits_{n=0} ^{+\inf} a_n A_{n,k} I_{\l}(kr)\frac{\pd}{\pd r}I_{\l}(kr)\,, \mbox{if $r<r_i$}\,, \\
\frac{q^2}{4\pi} \frac{L_{\om_e}}{ [ \r(r) ] ^2} -\frac{q^2}{\om_e} \int \limits_{0}^{+\inf}dk \sum\limits_{n=0}^{+\inf} a_n B_{n,k}K_{\n}(k\r(r))\frac{\pd}{\pd r}K_{\n}(k\r(r))\,,  \mbox{if $r>r_i$}\, ,
\ea \right.
\ee
where we use the global coordinate $r$ (defined in Sect. \ref{s2}) to represent the results with a numerical evaluation by plotting the self-force as a function of the continuous quotient $r/r_i$. In Figs. \ref{1} and \ref{2} we show the self-fore on the four different types of possible backgrounds. Figure \ref{1} corresponds to examples of concave ($\kappa<0$) shells made of ordinary matter, while Fig. \ref{2} to convex ($\kappa>0$) shells constructed with exotic matter. Positive values of the self-force correspond to a repulsive force from the central axis of symmetry of the space-time. 

Figure \ref{1a} represents cases where $\om_i=1>\om_e$; the self-force vanishes at the axis as expected (there is cylindrical symmetry and no singularity at $r=0$) and, asymptotically far, Linet's term is manifest due to the global deficit angle appearing beyond $r=r_i$. In terms of the equivalent problem, in the interior region $\Om_{int}$ the renormalized field has a vanishing regularized bulk term (Minkowski case) and the only contribution arises from the shell term $\P^{\s_1}$. In the exterior region $\Om_{ext}$, both $\P^{\s_2}$ and ${\lbr \P^{bulk}_2 \rbr}_R=\P^{Linet}_{\om_e}$ are present because $\om_e \neq 1$. We see in either region that near $r=r_i$ the charge is repelled from the concave ($\kappa<0$) shell of ordinary matter. We associate this repulsion with a surface charge density $\s_{1(2)}$ in the equivalent problems from each side of the shell. For the exterior problem, the asymptotic behavior coincides with the cosmic string repulsion $\P^{Linet}_{\om_e}$, we shall see that $\P^{\s_2}$ is subleading. In Fig. \ref{1b}, $1>\om_i>\om_e$; $\kappa<0$ with ordinary matter density and the charge is repelled from the shell again. The difference is the diverging repulsion from the straight cosmic string at $r=0$ coming from $\lbr \P^{bulk}_1 \rbr_R=\P^{Linet}_{\om_i}$. As a consequence there is a stable equilibrium position at some $r \in (0;r_i)$.

Figure \ref{2a} has $\om_i<\om_e=1$; the shell is made of exotic matter, with $\kappa>0$, and the charge is attracted toward it. In the interior region the repulsive effect from Linet's regularized bulk term is enforced by an attraction to the shell. At the other side of the shell space-time is Minkowski and the equivalent problem has a unique contribution coming from the shell's field, which attracts the charge. Finally, Fig. \ref{2b} shows the case $\om_i<\om_e<1$, which manifests the same qualitative behavior of the latter up to the vicinities of the shell. Sufficiently far from $r=r_i$ Linet's monopolar repulsive force from the centre is manifest and, as a consequence, an unstable equilibrium point appears at some $r>r_i$.

\begin{figure} [h!]
 \centering
  \subfloat[$\s_2$ for a shell with $\kappa<0$; $\om_i=1$,
$\om_e=0.9$ (ordinary matter).]{
   \label{3a}
    \includegraphics[width=0.35\textwidth]{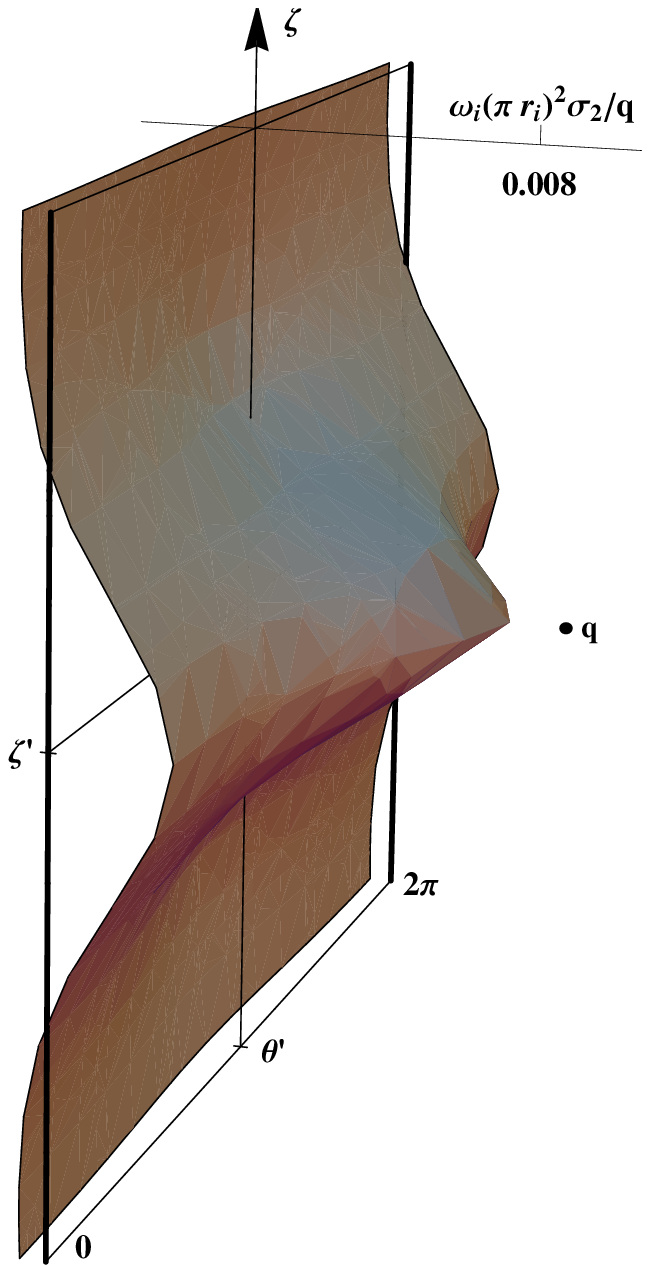}}
  \qquad \subfloat[$\s_2$ for a shell with $\kappa>0$;
$\om_i=0.5$, $\om_e=1$ (exotic matter).]{
   \label{3b}
    \includegraphics[width=0.3\textwidth]{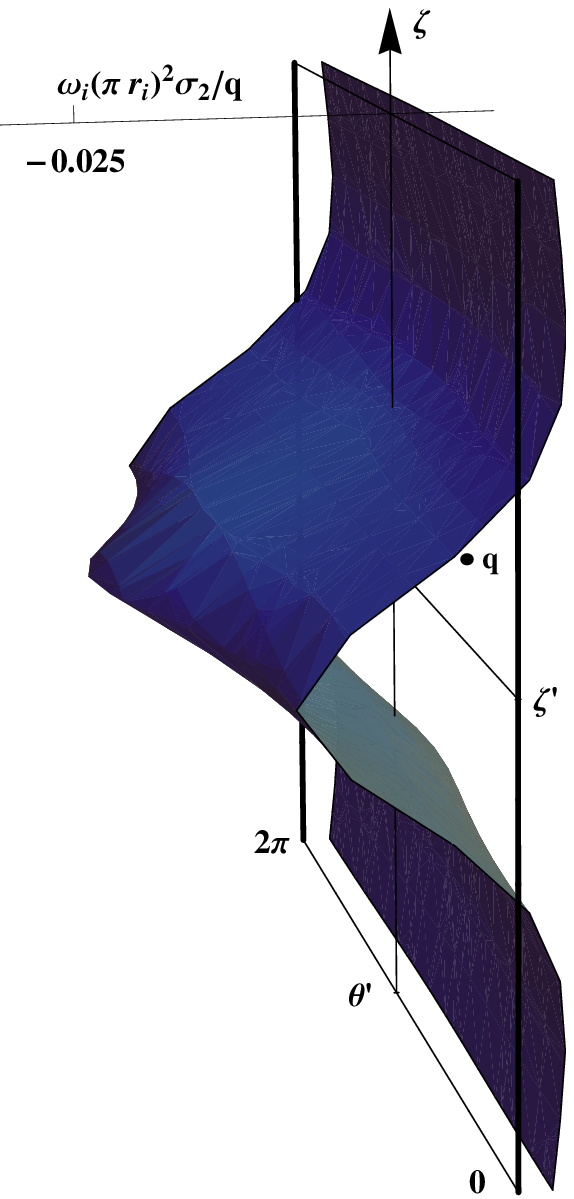}}
 \caption{Three-dimensional representation of the surface charge density $\s_2$. The vertical plane represents the two-dimensional spatial surface at $\pd\Om_{ext}$ in coordinates $(\t,\zeta=z/r_i)$ for the equivalent problem in the exterior region. Positive values of $\s_2(\t,\zeta) \frac{\om_i(r_i \pi)^2}{q}$ appear to the right of the plane in both cases. The charge, schematically drawn, is placed at $(\t'=\pi,z')$ and $r'=2r_i$ in the respective space-times of these examples.}
 \label{3}
\end{figure}

From (\ref{sig}) we compute explicitly $\s_a$ for the equivalent problems in consideration and analyze how the shell affects the charge $q$ depending on their relative distance and the relation with the extrinsic curvature:
\be \ba{ll} \lb{sigtsg}
\s_a(\t,z)= \int \limits_{0}^{+\inf}dk\, Z(k,z) \sum \limits_{n=0}^{+\inf} Q_n(\t)\, \s_{n,k}^{(a)}\,, \\
\s_{n,k}^{(a)} = \frac{q}{4 \pi} s_{n,k}^{(a)} 
\left\{\ba{ll}
\frac{1}{r_i} \,,  \quad \mbox{if $a=1$}\,, \\
\frac{1}{r_e} \,,  \quad \mbox{if $a=2$}\, .
\ea \right. 
\ea  
\ee
We point out that these $\s$ are a construction for the equivalent problems and do not appear in the original one where Eq. (\ref{jump}) holds in general. Note that Fourier coefficients $\s_{n,k}$ depend on the position $r'$ of the charge, see (\ref{potsig}), as image sources do in usual electrostatic problems. The net total charge $Q_a$ in the surface with $\s_a$ is in general:
\be \lb{Q}
Q_a = \sqrt{-g(r_i)} \int\limits_{-\inf}^{+\inf} dz \int_{0}^{2 \pi} d\t \, \s_a(\t,z) = 4 \pi \sqrt{-g(r_i)} \lim_{k \to 0} \s_{0,k}^{(a)}\,.
\ee
In our problems the explicit calculation from (\ref{sigtsg}) shows
\be
\frac{Q_a}{q}= \left\{ \ba{ll}
\frac{\om_i-\om_e}{\om_e}\quad \mbox{if} \,\, a=1\,, \\
\quad 0 \qquad \mbox{if} \,\, a=2\,.
\ea \right.
\ee
A clear geometric quantity is found in the net charge of the interior equivalent problem: 
\be \lb{q1}
\frac{Q_1}{q} = - \kappa \, r_i \, \frac{\om_i}{\om_e} = - \kappa \, r_e \,.
\ee
We find that the total \textit{effective shell charge} from the inside region is proportional to $\kappa$ (the trace of the extrinsic curvature jump) and the sign of $-Q_1/q$ is exclusively determined from the concavity of the shell, moreover, it does not depend on the position of $q$. A numerical analysis of (\ref{sigtsg}) shows $Q_1$ is distributed with a central peak at $(\t',z')$ and that the quotient $\s_1(\t,z)/q$ has the same sign as $Q_1/q$ all over the shell's layer density. 
On the other hand, for the exterior problem, $\s_2$ has null net charge as expected, $q$ is the total charge asymptotically. Despite this, $\s_2(\t,z)$ has a peak with definite net charge in a neighborhood of $(\t',z')$; for this case we show some examples plotted in Fig. \ref{3}, a three-dimensional representation of $\s_2$ over the surface $r_2=r_e$ near the position $(\t'= \pi, z'=\zeta' \, r_i)$. Two cases are considered; Fig. \ref{3a} corresponds to a concave shell of ordinary matter as in Fig. \ref{1a}, Fig. \ref{3b} corresponds to a convex shell of exotic matter as in Fig. \ref{2a}. The vertical plane in the picture represents a portion of the cylindrical surface, $\t=0$ and $\t=2 \pi$ have to be identified. The density distributions to the right of this plane represent positive values of the quotient $\s_2/q$. In both cases we see a peak centered at $(\t',z')$. This peak has the same sign as $q$ if $\kappa<0$ and an opposite sign if $\kappa>0$, so this \textit{image} charge concentration at the centre of the shell is responsible for the local behavior, repulsive or attractive respectively, of the self-force near the shell for the exterior problem. 

In general, the function $\s_a(\t,z)$ is well defined by construction except in the limit $r' \rightarrow r_i$; for our cases we find
\be
\lim_{r' \to r_i} \frac{\s_a(\t',z')}{q \, r_i^2} \rightarrow \left\{ \ba{ll}
+ \inf\,, \quad \mbox{if $\kappa<0$ (ordinary matter)}\,, \\
- \inf\,, \quad \mbox{if $\kappa>0$ (exotic matter)}\,,
\ea \right.
\ee
which is a final check for the relation between the extrinsic curvature and the effective charge density $\s_a$ of the \textit{shell} field from either side. Returning to the original problem in the locally flat space-time with a thin-shell of matter, we can give a geometric summary of the results as follows: a convex ($\kappa>0$) or concave ($\kappa<0$) time-like shell $\pd \Omega = \{ r=r_i \}$ focuses or defocuses the electrostatic field lines of a charge $q$ contributing with an attractive or repulsive effect on the self-force, respectively.

\subsection{Complementary example: thin-shell wormholes}
\lb{s5.3}

As mentioned previously in the introduction, thin-shell geometries are vastly used in the problem of self-forces, specially in wormhole space-times. These geometries are generally constructed with exotic matter at the shell of the throat and $\kappa>0$, so we expect that in cylindrical wormholes with static diagonal metrics the corresponding \textit{shell field} contributes with an attractive (focusing) electric force toward the throat. Moreover, motivated by result of Eq.  (\ref{q1}), we may ask if there is a similar relation between the \textit{effective image charge} $Q^{wh}$ and the concavity of the throat for the  equivalent problems stated in wormholes. We point out that for thin-shell wormholes (TSWH) the electric field flux as measured from one asymptotic region is not fixed in $q$ as it is when there is an interior complementary region. The equivalent problems in the bulk where $q$ is placed always admit a \textit{shell field} sourced by some $\s_a$ with net effective total charge, because beyond the throat there is another asymptotic region. Applying the equivalent problem interpretation for a test charge in a locally flat cylindrical TSWH with conical asymptotics (similar to those considered in \cite{eoc}) we obtain
\be
\frac{Q_-^{wh}}{q}= - \frac{\om_+}{\om_- + \om_+} = - \frac{1}{\kappa_{wh}\, r_+}\,,
\ee
with $q$ placed in region $\Om_-$, $2\pi(1-\om_{\pm})$ the deficit angles at each side of the wormhole throat, and $\om_- r_- = \om_+ r_+$ given from the junction condition (some details for this calculation are specified in the appendix at the end). In the wormhole case the asymptotic behavior of the \textit{shell} field $\P^{\s_{\pm}}$ is inversely proportional to the concavity $\kappa_{wh}$. This is an amazing result in probing global aspects of wormhole geometries with electrostatics; in this case the monopolar term of $\P^{\s_{\pm}}$ is a measure of the concavity of the throat.

\section{Summary}
\lb{s6}

The electrostatic self-force on a point charge in cylindrical thin-shell space-times is interpreted as the sum of a \textit{bulk} field and a \textit{shell} field. To this end we developed the equations for the electrostatic field potential using separation of variables in cylindrical coordinates. We defined the \textit{bulk} term so that it corresponds to a field sourced by the test charge placed in a space-time without the shell. This field contains the divergent part of the potential to be renormalized with known procedures \cite{DW,kpl}. The \textit{shell} field left over accounts for the discontinuity of the extrinsic curvature ${\kappa^p}_q$ of the TSST and contains information of the two constituent geometries of the complete space-time. We use this formal development to define an equivalent electrostatic problem in which the geometrical effect of the shell of matter over the test charge $q$ is replicated with the electric potential produced by a non-gravitating surface charge density $\s$ of net total \textit{image} charge $Q$. With this procedure we analyzed the \textit{shell} field in the interior and exterior regions of a TSST and motivate a new way of measuring and interpreting some global properties of space-time with electrostatics. 

Throughout this work, the electrostatic problem of a point charge $q$ in a locally flat geometry with a cylindrical thin-shell of matter is solved explicitly to apply the interpretation method suggested. The self-force obtained is radial. The regularized \textit{bulk} field, known as the Linet cosmic string repulsive term, acts as a monopolar electric field from the central axis of the space-time. The analysis of the \textit{shell} field shows that it contributes with a repulsive electric force from the shell if $\kappa<0$ (ordinary matter) and an attractive electric force toward the shell if $\kappa>0$ (exotic matter). We find that the total image charge is zero for exterior problems (charge placed in the asymptotic region), while for interior problems $Q/q=-\kappa \, r_e$, with $r_e$ the external radius of the shell. In the latter case we see that a quantitative measure of the concavity $\kappa$ of the thin-shell is obtained from the monopolar \textit{shell} field. These results are new and offer a better understanding of the self-force problem in space-times with shells. The procedure can be applied to interpret self-forces in other space-times with shells, and the above result motivated the same analysis for wormhole space-times. We took a previous work on self-forces on a charge in locally flat cylindrical TSWH \cite{eoc}, and we calculated the total net \textit{effective charge} of the shell and found an inverse proportional relation with the concavity of the throat: $Q^{wh}_{\mp}/q=-1/ (\kappa_{wh} r_{\pm})$. We may ask if similar results appear in other wormholes, or in spherical TSST, relating \textit{shell} fields and the extrinsic curvature discontinuity. We think this open question could motivate future work in the area.

\section*{Acknowledgments}
\lb{s7}
I want to thank C. Simeone for his expert suggestions. This work was supported by the National Scientific and Technical Research Council of Argentina.

\section*{Appendix}
Following the same procedure applied for cylindrical TSST developed in the main body of this work, we reproduce similar results to \cite{eoc} but with an asymmetric locally flat TSWH. This is constructed with the exterior regions $\Om_{\mp}=\{ x_{1,2}:\, r_{1,2} \in \lbr r_{\mp},+\inf \rbr \}$ of two gauge cosmic string geometries of deficit angles $2\pi(1-\om_{\mp})$, respectively. The junction condition establishes $r_-\,\om_-=r_+\,\om_+$. The difference with the previous case is that in TSWH we have two asymptotic regions and condition (\ref{0}) is replaced with (\ref{inf}) for both radial functions. Additionally, in Eq. (\ref{jump}) the minus sign in the second term is changed for a plus sign. The procedure can be applied analogously step by step using the coordinate system of patches $x_1$ and $x_2$, which cover the complete manifold $\Om = \Om_- \cup \Om_+$. The regularized electrostatic potential for a charge $q$ at rest in $(r'_{1(2)},\t',z')$ in either region is
\be
\P_R= \P^{bulk}_R +\P^{\s}\,.
\ee
In the wormhole space-time, the equivalent problems in each side of the throat have
\be\lb{bulkwh}
\P^{bulk} = \P^{CS}_{\om_{\mp}}(r_{1,2},\t,z)\,,
\ee
for $x_{1,2}, x_{1,2}' \in \Om_{\mp}$, respectively, and
\be\lb{potsigwh} \ba{ll}
\P^{\s_{\mp}} = q \int \limits_{0}^{+\inf}dk\, Z(k,z) \sum\limits_{n=0}^{+\inf} Q_n(\t) s_{n,k}^{\mp} 
I_{\u_{\mp}}(k r_<) K_{\u_{\mp}}(k r_>)\,,  \\\\

s_{n,k}^{\mp} = 
\frac{W^{\mp}_{n,k}}{\om_{\mp}} \frac{K_{\u_{\mp}}(k r'_{1,2})}{I_{\u_{\mp}}(k r_{\mp})}\,, \quad \u_{\mp}=n/\om_{\mp} \,,\\\\

W^{\mp}_{n,k} = -\frac{I_{\u_{\mp}}(k r_{\mp}) \lbr \pd_r K_{\u_{\pm}}(kr) \rbr |_{r_{\pm}}  + K_{\u_{\pm}}(k r_{\pm}) \lbr \pd_r I_{\u_{\mp}}(k r) \rbr |_{r_{\mp}}}{K_{\u_{\pm}}(k r_{\pm}) \lbr \pd_r K_{\u_{\mp}}(k r) \rbr |_{r_{\mp}} + K_{\u_{\mp}}(k r_{\mp}) \lbr \pd_r K_{\u_{\pm}}(k r) \rbr |_{r_{\pm}}} \,,
\ea  
\ee
where $r_{\stackrel{>}{<}}=\left\{ r_{1,2}; r_{\mp} \right\}$. The surface charge density of the equivalent problems are
\be \ba{ll} \lb{sigwh}
\s_{\mp}(\t,z) = \int \limits_{0}^{+\inf} dk\, Z(k,z) \sum\limits_{n=0}^{+\inf} Q_n(\t)\, \s_{n,k}^{\mp} \,,\\
\s_{n,k}^{\mp} = \frac{q}{4 \pi} \frac{s_{n,k}^{\mp}}{r_{\mp}} \,. 
\ea  
\ee
Finally from Eq. (\ref{Q}) we obtain the total net image charge
\be 
\frac{Q^{wh}_{\mp}}{q}=-\frac{\om_{\pm}}{\om_{-}+\om_{+}} = -\frac{1}{\kappa_{wh}} \frac{\om_{\pm}}{\om_{\mp}\,r_{\mp}}= -\frac{1}{\kappa_{wh}\,r_{\pm}} \,.
\ee

\end{document}